# Numerical simulation of nano scanning in intermittent-contact mode AFM under Q control

**A Varol, I Gunev, B Orun, and C Basdogan**[1]

College of Engineering, Koc University, Istanbul, Turkey, 34450

E-mail: cbasdogan@ku.edu.tr

**Abstract**

We investigate nano scanning in tapping mode atomic force microscopy (AFM) under quality (Q) control via numerical simulations performed in SIMULINK. We focus on the simulation of whole scan process rather than the simulation of cantilever dynamics and the force interactions between the probe tip and the surface alone, as in most of the earlier numerical studies. This enables us to quantify the scan performance under Q control for different scan settings. Using the numerical simulations, we first investigate the effect of elastic modulus of sample (relative to the substrate surface) and probe stiffness on the scan results. Our numerical simulations show that scanning in attractive regime using soft cantilevers with high $Q_{eff}$ results in a better image quality. We, then demonstrate the trade-off in setting the effective Q factor ($Q_{eff}$) of the probe in Q control: low values of $Q_{eff}$ cause an increase in tapping forces while higher ones limit the maximum achievable scan speed due to the slow response of the cantilever to the rapid changes in surface profile. Finally, we show that it is possible to achieve higher scan speeds without causing an increase in the tapping forces using adaptive Q control (AQC), in which the Q factor of the probe is changed instantaneously depending on the magnitude of the error signal in oscillation amplitude. The scan performance of AQC is quantitatively compared to that of standard Q control using iso-error curves obtained from numerical simulations first and then the results are validated through scan experiments performed using a physical set-up.

[1] Corresponding Author

## 1. Introduction

In tapping-mode dynamic force microscopy (DFM), a cantilever probe, oscillating in free-air around resonant frequency with amplitude of $A_0$, is used to scan a sample surface [1]. When the tip of the probe taps the sample surface lightly for a very short period of time, the oscillation amplitude is reduced to $A<A_0$. In amplitude modulation scheme, a scan controller moves the sample or the probe in vertical direction (i.e. z-direction) such that the oscillation amplitude during tapping stays constant at set amplitude $A_{set}$. These up and down movements in vertical direction are recorded during scanning to construct the surface profile of the sample.

During tapping-mode scanning, the interaction forces between the probe tip and the sample surface are highly nonlinear and the response of the cantilever probe to these forces is primarily governed by its quality factor, $Q = \Delta\omega/\omega_r$, where $\omega_r$ is the resonance frequency of the cantilever and $\Delta\omega$ is the width of the resonance curve for which the energy is greater than the half of its peak. The Q factor of a cantilever probe indicates its energy dissipation capacity or damping present in the system. A probe with low Q factor dumps its energy faster, resulting in lower-amplitude steady-state oscillations and a rounded resonance curve. On the other hand, a probe with high Q factor oscillates more and its resonance curve shows a sharp peak.

The Q factor of a cantilever probe can be set to a value using an additional feedback circuit. By applying an appropriate force to the oscillating cantilever, its motion can be regulated in such a way that the modified response of the system shows an increased or decreased Q factor. This approach is known as "Q control" and has been suggested as an effective method in DFM [2]. In figure 1, a schematic illustration of a Q controlled cantilever probe is presented. Typically, a phase shifter and an amplifier with a gain G are used in the feedback circuit in order to control the Q factor of an oscillating cantilever. First, the displacement signal of an oscillating probe is measured using a photo-detector, shifted in phase using the phase shifter, and then scaled by the gain G, and finally used as the velocity signal in the feedback loop. This velocity signal is added to (or subtracted from) the driving signal to decrease (or to increase) the effective damping of the cantilever. We hereafter will call this approach where the gain G is set to a constant value before the scan process as standard Q control in the text.

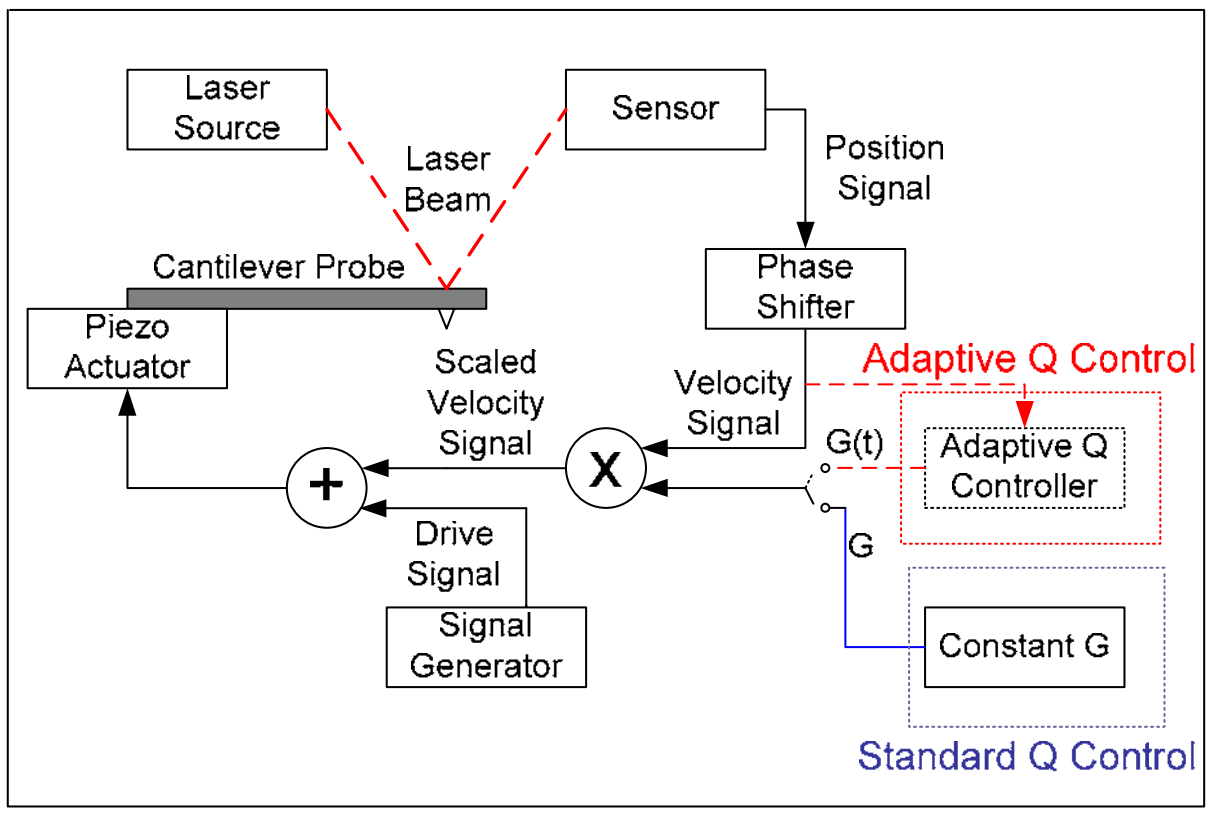

**Figure 1** A schematic representation of driven cantilever probe in tapping mode scanning under standard (blue solid line) and adaptive (red dashed line) Q control. The gain G is constant in standard Q control whereas it is adjusted in real-time based on the surface profile in adaptive Q control.

Studies have shown that the image resolution improves with Q control, but the mechanism for this is not well-understood [3-5]. Rodriguez and Garcia developed an analytical solution for an oscillating probe under Q control [6]. Their simulations suggested that the slope of the amplitude-distance curves is substantially larger implying a higher sensitivity to tip-surface variations, which results in better image quality. However, numerical simulations performed later by Kokavecz et al yielded that the slope of the amplitude versus distance curves is not increased for stiff samples and hence they concluded that the maximal probe sensitivity cannot be increased with Q control [7]. Holscher et al performed numerical simulations and showed that an increased Q factor prevents the oscillating cantilever to jump into a repulsive imaging regime during tip-sample approach, which often occurs in tapping mode imaging without Q control [8]. They concluded that restriction of the maximal tip-sample force to specific parts of the attractive regime in Q control is the main reason for the enhanced imaging quality. Hence, it can be argued that Q control keeps the probe in the attractive regime longer and as a result, the magnitude of the average tapping forces is reduced. In fact, Jäggi et al experimentally determined that the average tip-sample forces are reduced by Q control [9].

On the other hand, there is a trade-off between tapping forces and scan speed when setting the Q factor of a scanning probe, which has not been investigated in detail. For example, when scanning in liquid, the Q factor of the probe is increased to obtain better scan results and prevent the damage on the sample [10]. On the other hand, the mechanical sensing bandwidth of the probe is inversely proportional to its Q factor, and increasing it limits the maximum achievable scan speed (i.e. the maximum speed that the probe can trace the sample surface with a reasonable amount of positional error in scan profile). Sulchek et al showed that the sensing bandwidth of a scanning probe and scan speed can be improved significantly by actively lowering the Q factor of the probe when scanning nano surfaces in air [11, 12].

In standard Q control, the effective Q factor ($Q_{eff}$) of the probe is set to a value that is lower or higher than its native one before scanning and achieving higher scan speeds with reduced tapping forces is not possible. Gunev et al suggested that these two benefits can be realized simultaneously using adaptive Q control (AQC) [13]. They developed a signal processing circuit that adjusts the gain G on the fly during scanning depending on the error in amplitude signal. If there is a tendency towards saturation in error (as it typically occurs in scanning steep downward steps), the controller increases the Q factor of the probe rapidly to avoid the problem.

In this study, we investigate nano scanning in tapping mode atomic force microscopy (AFM) under quality (Q) control via numerical simulations. We first explore the influence of elastic modulus of sample material (relative to the substrate surface) and probe stiffness on scan performance. We then show the trade-off between scan speed and tapping forces in setting the Q factor of the probe. Finally, we demonstrate that AQC solves the trade-off problem and expands the allowable workspace of the scan controller. In most of the earlier numerical studies, only the cantilever dynamics has been investigated, but the transient effects during scanning have been neglected (i.e. the steady state solution of cantilever oscillations is considered only). These studies have primarily investigated the effect of various parameters and tip-sample forces on the cantilever dynamics. Moreover, the parameters of the cantilever model used in these simulations are mostly selected somehow arbitrarily to meet the simulation needs and not obtained from the experimental measurements. We categorize this type of numerical studies as “cantilever” simulations to differentiate it from our work of “scanning” simulations. To our knowledge, there are only a few studies focusing on the end-to-end simulation of the whole scan process [14]. Our simulations not only integrate the model of cantilever, but also the models of other components of the scanning system such as the scan controller, measurement devices, and DAQ sampling unit. This is achieved using SIMULINK, which provides a flexible computing environment for supporting linear and nonlinear systems, modeled in continuous time, sampled time, or a hybrid of the two. The detailed models of the components used in our simulations are discussed in the upcoming sections. Moreover, our numerical simulations are supported by the experiments conducted using a home-made AFM set-up.

## 2. Set-up

The major components of our AFM set-up include a self-actuated AFM probe which is brought close to a sample surface using an XYZ manual stage, a computer controlled XYZ nano-stage for moving the sample surface with respect to the probe, and a Laser Doppler Vibrometer (LDV) for measuring the vertical vibrations of the probe. An analog signal processing circuit consisting of a) a root mean square (RMS) converter, b) a variable phase-shifter, and c) a voltage-multiplier is built and integrated into the AFM set-up to adaptively modify the Q factor of the probe on the fly during scanning. The details of the set-up and the components are available in our earlier publication [13].

In order to scan a sample using the developed AFM system, the velocity of the probe tip is measured as a continuous signal using the LDV and then the RMS value of this signal is calculated over a running window and sampled by a DAQ card into a computer. The oscillation amplitude of the probe A is calculated from the RMS signal and then compared with the desired oscillation amplitude $A_{set}$. A proportional integral (PI) controller developed in LabVIEW keeps the vibration amplitude of the probe constant during raster scanning by moving the nano stage up and down along the Z-axis based on the error signal ($A_{set}$ - A).

## 3. Cantilever Dynamics

### *3.1. Cantilever Model*

A damped mass-spring system is used to model the dynamical behavior of the cantilever. This model is typically utilized by the earlier AFM studies and it is a reasonable approximation of the oscillating cantilever [10, 15, 16]. It is assumed that the cantilever is externally driven by a sinusoidal force $F_{drive}$ at its resonance frequency ω. In addition to the driving force, the oscillations of the cantilever are also influenced by the interaction forces between the cantilever tip and sample surface, $F_{ts}$. This force is a function of tip-sample separation distance h. The dynamics of the cantilever probe can be written as a second-order differential equation in the form of

$$m\ddot{z} + b\dot{z} + kz = F_{drive} + F_{ts}(h) \tag{1}$$

where, k and m are the effective spring constant and mass of the cantilever respectively. In addition, a damper with a coefficient b is added to the model to simulate damping due to air. The damper applies a resistance to the oscillations proportional to the vibration velocity of the cantilever and causes energy dissipation. Expressing the above equation in terms of measurable quantities, we have the following equation

$$m\ddot{z} + \frac{m\omega}{Q}\dot{z} + kz = F_0 \cos(\omega t) + F_{ts}(h) \tag{2}$$

where, ω and $F_0$ are the frequency and the magnitude of the external driving force, respectively (the cantilever probe is actuated at its resonance frequency, $\omega = \omega_r$). Q factor of the cantilever is inversely proportional with the damping coefficient b. For a system with high Q factor (low damping), the resonance frequency of the system is approximately equal to its natural frequency, $\omega_r \approx \omega_n = \sqrt{k/m}$.

The Q factor and the resonance frequency ω of the cantilever can be determined experimentally by examining its frequency response. The frequency response of the cantilever used in our experimental set-up is plotted for different gains G and matched to the one used in our numerical simulations (figure 2). From the plots, the resonance frequency of the cantilever (260 kHz) is determined as the frequency at which the oscillation amplitude reaches to maximum and the Q factor of the probe ($Q_{native}$ = 311) is determined  by measuring the frequency range Δω where the energy of the oscillations is greater than half of the maximum energy at the resonance frequency. The value of the spring constant used in the numerical model is obtained from the catalogue of the manufacturer of the cantilever (the nominal value is reported as 3 N/m at the operating frequency).

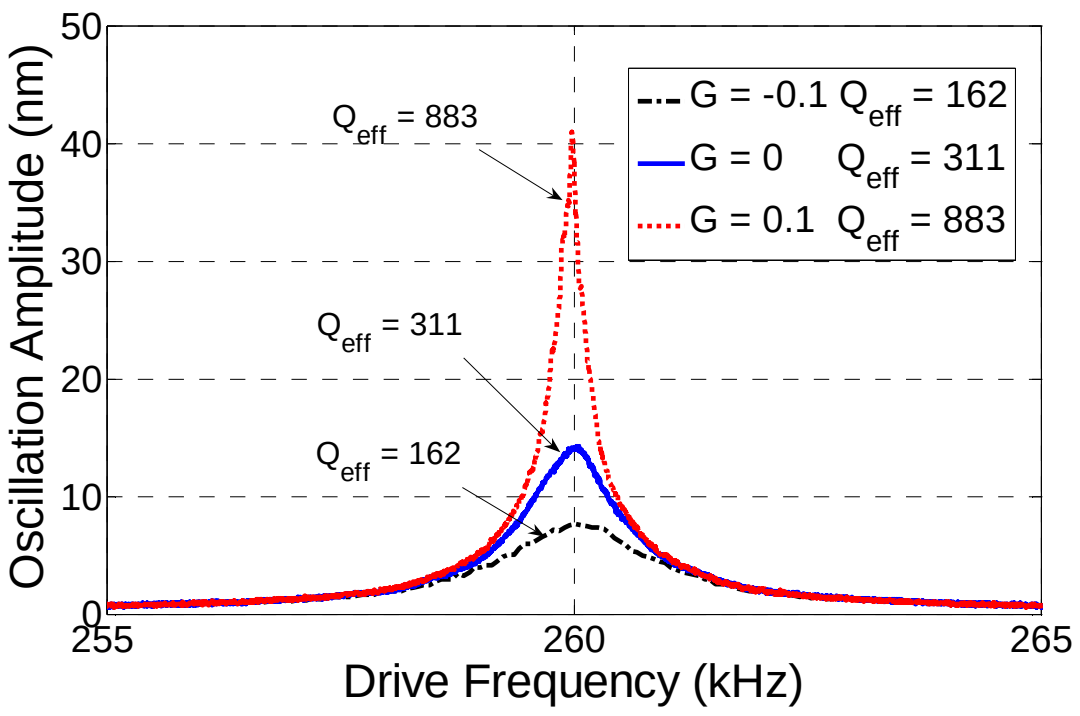

**Figure 2** The amplitude response of the cantilever around the resonance frequency for different values of effective quality factor is obtained through experimental measurements.

*3.2. Force Interactions*

The force interactions between the probe tip and sample geometry is modeled as a spherical object interacting with a flat surface [17-19]. In the tapping mode AFM, the cantilever is oscillated over the sample surface and contacts the surface for a brief period of time at each oscillation cycle. As a result, the distance h between the cantilever tip and sample surface changes continuously (figure 3a). Depending on the separation distance, two different interaction models are typically utilized to calculate the tip-sample forces.

*1) Non-contact Forces*: For the separation distances larger than the inter-atomic distance $a_0$, the long range attractive force is dominant (note that electro-static interactions are ignored). The attraction results from the integration of van der Waals energy between two atoms over the atoms of interacting surfaces. The van der Waals attraction force between a sphere and a flat surface is the negative gradient of this energy and can be written as a function of tip-sample separation distance h, the tip radius R, and the Hamaker constant H as follows [20]

$$F_{vdw}(h) = \frac{HR}{6h^2} \qquad \text{if } h > a_0. \qquad (3)$$

*2) Contact Forces*: It is assumed that a mechanical contact between the cantilever tip and the sample surface occurs when the separation distance h is smaller than the inter-atomic distance $a_0$ (negative values of h indicate indentation into the sample surface). During the contact, both adhesive and repulsive forces are effective. According to the DMT theory [21], the adhesion force is equal to the van der Waals attraction force when h = $a_0$. The repulsive force arising from the mechanical contact of the tip with the sample surface is modeled using contact mechanics. Hence, the total force acting on the cantilever tip due the adhesive and repulsive components is given by

$$F_{contact}(h) = -\frac{HR}{6a_0^2} + \frac{4}{3}E^*\sqrt{R}\left(a_0 - h\right)^{3/2} \qquad \text{if } h < a_0. \qquad (4)$$

$$\frac{1}{E^*} = \frac{(1-\nu_t)^2}{E_t} + \frac{(1-\nu_s)^2}{E_s}$$

where E* is the effective Young's modulus of the tip-sample pair, $E_t$ and $E_s$ are the elastic moduli of the tip and surface materials, $\nu_t$ and $\nu_s$ are the Poisson’s ratio of the tip and the surface material, respectively.

Combining the contact and non-contact forces, the interaction force $F_{ts}$ between the tip and the surface can be expressed as:

$$F_{ts}(h)=\begin{cases}-\dfrac{HR}{6h^2} & \text{if} \quad h>a_0 \\ -\dfrac{HR}{6a_0^2}+\dfrac{4}{3}E^*\sqrt{R}\left(a_0-h\right)^{3/2} & \text{else}\end{cases} \tag{5}$$

## 4. SIMULINK Model

Simulating the dynamical behavior of the cantilever alone does not help us to investigate the scan performance under Q control for different scan settings. We also developed the models of the individual scan components and integrated them with the model of the cantilever to perform end-to-end scanning simulations. In addition to the force and the cantilever models described in the previous section, the complete model of the scanning system (shown in figure 3b) includes the numerical models of the following physical components:

a) *Vibrometer*: The Laser Doppler Vibrometer (LDV) in the physical set-up is modeled as a block which differentiates the vibration signal (i.e. AC deflection of the cantilever) first and then outputs it after adding a time-delay on it. This time-delay is caused by the digital signal processing unit of the LDV and confirmed to be fixed for the operating frequency of the cantilever [13].

b) *RMS Converter*: This block is used to compute the RMS of the vibration velocity. A built-in block (RMS in SimPowerMechanics Library) is used for this operation and the RMS value of the input signal is calculated over a running window. In order to calculate the oscillation amplitude of the cantilever, the RMS of the velocity signal is multiplied by a proper gain K as shown in figure 3b.

c) *PI Scan Controller*: This block is used to control the vertical movements of the computer-controlled XYZ stage. It has an analog to digital converter that samples the error signal at a fixed sampling rate. A built-in block (*pid* controller) is utilized to actuate the XYZ stage based on the error in the peak-to-peak amplitude of the cantilever oscillations ($A_{set}$ - A).

d) *XYZ Stage*: In order to model the dynamics of the computer controlled XYZ nano stage used in the physical set-up, we have used a first order transfer function with a time constant smaller than the sampling time of the PI controller. The vertical movements of the XYZ stage are represented as $s_2$ in figure 3.

e) *Adaptive Q Controller*: This block is used to calculate the gain G(t) which depends on the peak-to-peak oscillation amplitude of the cantilever. If G(t) is set to a constant value, this block functions as a standard Q controller. In AQC, the gain is adjusted on the fly depending on the RMS value. The output of this block, G(t) is used to scale the velocity signal first and then the scaled signal ($F_{AQC}$) is added on the drive signal in order to change the $Q_{eff}$ of the cantilever.

In addition to the models of above components, the following blocks are used for regulating input-output operations:

i) *Input Profile*: This block generates the input surface profile as a function of time. In the case of scanning calibration steps with a constant height of $h_s$ and width of $w_s$, the input profile $s_1$ is periodic with a period of $t_s = w_s / v_s$, where $v_s$ is the scan speed. For the time interval $0 < t < t_s$, $s_1$ is given by

$$s_1 = \begin{cases} 0 & 0 < t < \frac{t_s}{2} \\ h_s & \frac{t_s}{2} < t < t_s \end{cases} \tag{6}$$

ii) *Output Profile*: This is the block where the inverse of the movements of the XYZ stage ($s_2$) are recorded. Hence, this is the output of the scanning system. The output scan profile should exactly match the desired input profile ($s_1$) under ideal conditions.

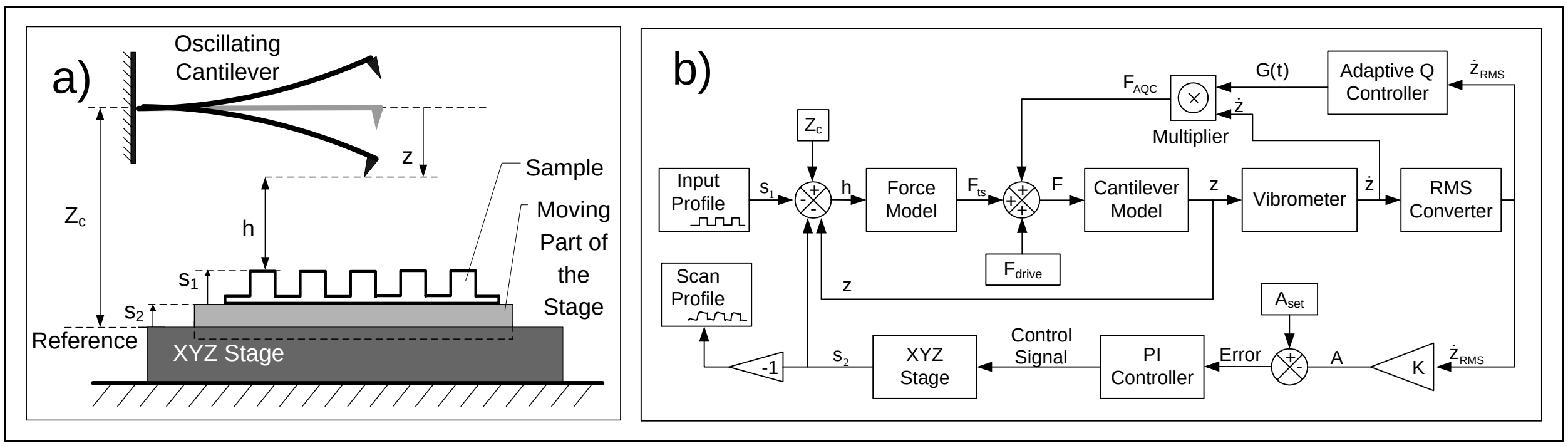

**Figure 3** SIMULINK model of our experimental scanning set-up.

In the model, the parameters $Z_c$, h, and z represent the vertical position of the resting cantilever with respect to the reference plane in the absence of tip-sample interaction, the vertical separation distance between the cantilever tip and the sample surface, and the instantaneous vertical position of the oscillating cantilever with respect to its resting position, respectively (figure 3a).

## 5. Scanning Simulations

In this section, we investigate AFM scanning under standard Q control first and then extend our work to show the benefits of AQC through numerical simulations. The numerical results for AQC are also validated by scanning experiments performed in our physical AFM set-up. In standard Q control, the gain G is set to a constant value prior to scanning in order to increase or decrease the $Q_{eff}$ of the cantilever and is not altered during scanning. In AQC, the $Q_{eff}$ of the cantilever is adjusted in real time depending on the surface profile by setting the gain G adaptively on the fly.

We first investigate the effect of Q factor and stiffness of a cantilever probe on its sensitivity in scanning soft and stiff samples (relative to the substrate surface) through numerical simulations. For this purpose, we performed scanning simulations with a 10 nm height soft sample ($E_{sample}$ = 200 MPa) lying on a stiffer substrate ($E_{substrate}$= 10 GPa). This is a typical scenario encountered when scanning soft samples in liquid where the Q factor of the cantilever is significantly reduced due to the environmental damping. The simulations are repeated for three different cantilevers (k = 0.05 N/m, 0.5 N/m, and 5 N/m) for $Q_{eff}$ varying from 10 to 800. All other parameters in the numerical model such as the PI controller gains, the material properties of the cantilever tip, scan speed, resonance frequency and free-air amplitude of the cantilever are kept constant throughout the simulations. As shown in figure 4a, the apparent (measured) height of the scanned sample depends on $Q_{eff}$ and cantilever stiffness. As $Q_{eff}$ is increased and k is decreased, the apparent height approaches to the actual value. Our numerical simulations show that scanning in attractive regime using soft cantilevers with high $Q_{eff}$ results in a better image quality. The sudden jumps to the actual value shown in figure 4a for k = 0.05 N/m and 0.5 N/m occurs when the cantilever starts to operate at the attractive regime. In this regime, only the non-contact forces are effective. The probe is not tapping on the sample surface and oscillation amplitude is reduced only by the non-contact interactions [19]. Since the probe tip is not in physical contact with the sample surface during this period, there is no sample deformation and hence no difference between the measured and actual heights of the sample.

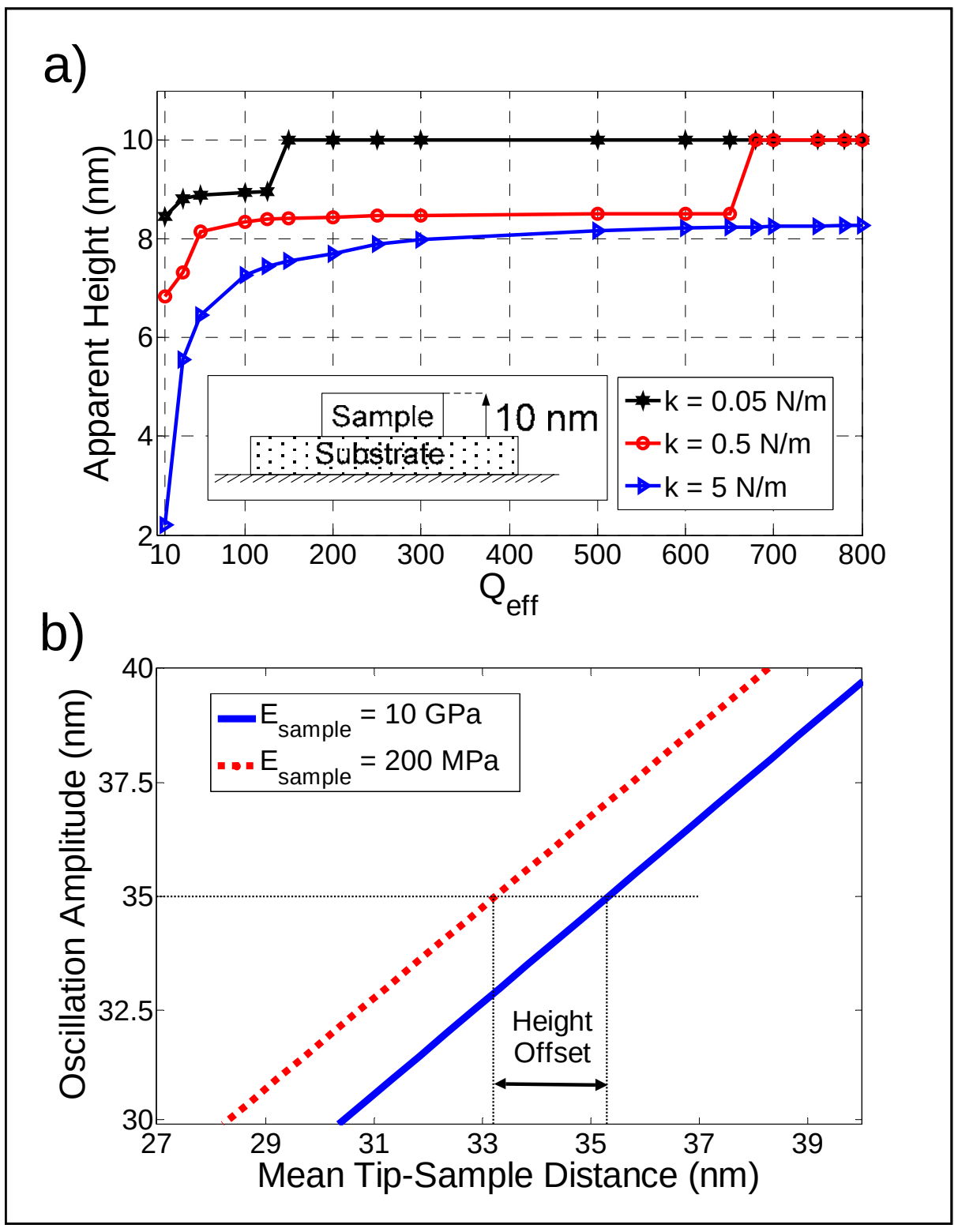

**Figure 4** (a) Apparent height of a 10 nm sample ($E_{sample}$ = 200 MPa) lying on a stiffer substrate ($E_{substrate}$ = 10 GPa) for different values of effective quality factor and stiffness constant of the cantilever. (b) Oscillation amplitude versus mean tip-sample distance for samples having different elastic modulus ($A_{set}$ = 35 nm, $A_0$ = 50 nm).

If the scanning is performed in the repulsive regime, where the cantilever tip touches the sample surface, the offset between the measured and actual heights is more prominent, especially for stiffer cantilevers. In this regime, the contact mechanics dominates the interactions between the probe tip and the sample surface. Hence, the elastic moduli of the cantilever tip and sample both influence the scan quality. In figure 4b, we show the change in the oscillation amplitude of a cantilever probe as a function of the tip-sample distance when the probe tip interacts with soft ($E_{sample}$ = 200 MPa) and stiff samples ($E_{sample}$ = 10 GPa) lying on a stiff substrate ($E_{substrate}$ = 10 GPa). The cantilever tip indents more into the soft sample than it does into the stiffer one to reach the same set amplitude and hence a height offset in the order of nanometers occurs when scanning soft samples. From the numerical simulations, we observe that this problem is more pronounced if the $Q_{eff}$ of the cantilever is low, as it occurs when the probe is immersed into liquid. These numerical simulations show a good agreement with the experimental studies of Ebeling et al [22] and Humphris et al [5]. In both of those studies, an increase in the apparent height of DNA on mica is reported when the effective quality factor of the cantilever is enhanced. In the case of scanning samples as stiff as the substrate surface, the height offset becomes insignificant. Although the scanning is performed in repulsive regime, the indentation of the probe tip into the sample surface is minimal throughout the sample surface independent of the cantilever stiffness.

Secondly, we consider the effect of Q factor on the tapping forces. In addition to the material properties of the sample, the force interactions between the probe tip and the sample surface depend on the free-air oscillation amplitude $A_0$ and the set amplitude $A_{set}$ of the probe when scanning under Q control. We performed numerical simulations to investigate the effect of $A_{set}$ (as a percentage of $A_0$) and $Q_{eff}$ on the tapping forces. In figure 5a, we present the magnitude of maximum tapping force as a function of $A_{set}/A_0$ for different values of $Q_{eff}$. The maximum tapping force is calculated based on the average of maximum indentations of the probe tip into the sample surface after the tapping amplitude reaches to steady state (i.e. $A = A_{set}$). As $A_{set}/A_0$ approaches to one, the probe tip starts tapping the sample surface lightly and the resultant interaction forces decrease. Similarly, as $Q_{eff}$ increases, the tapping forces also decrease. However, for a given $A_{set}/A_0$, the rate of decrease in tapping forces is less significant as $Q_{eff}$ increases (figure 5b). For scanning soft samples in liquid, keeping the tapping forces low is crucial in order to prevent damaging the sample. It is also desired to have lower tapping forces to prevent the tip wear during the scanning of stiffer surfaces in air as well.

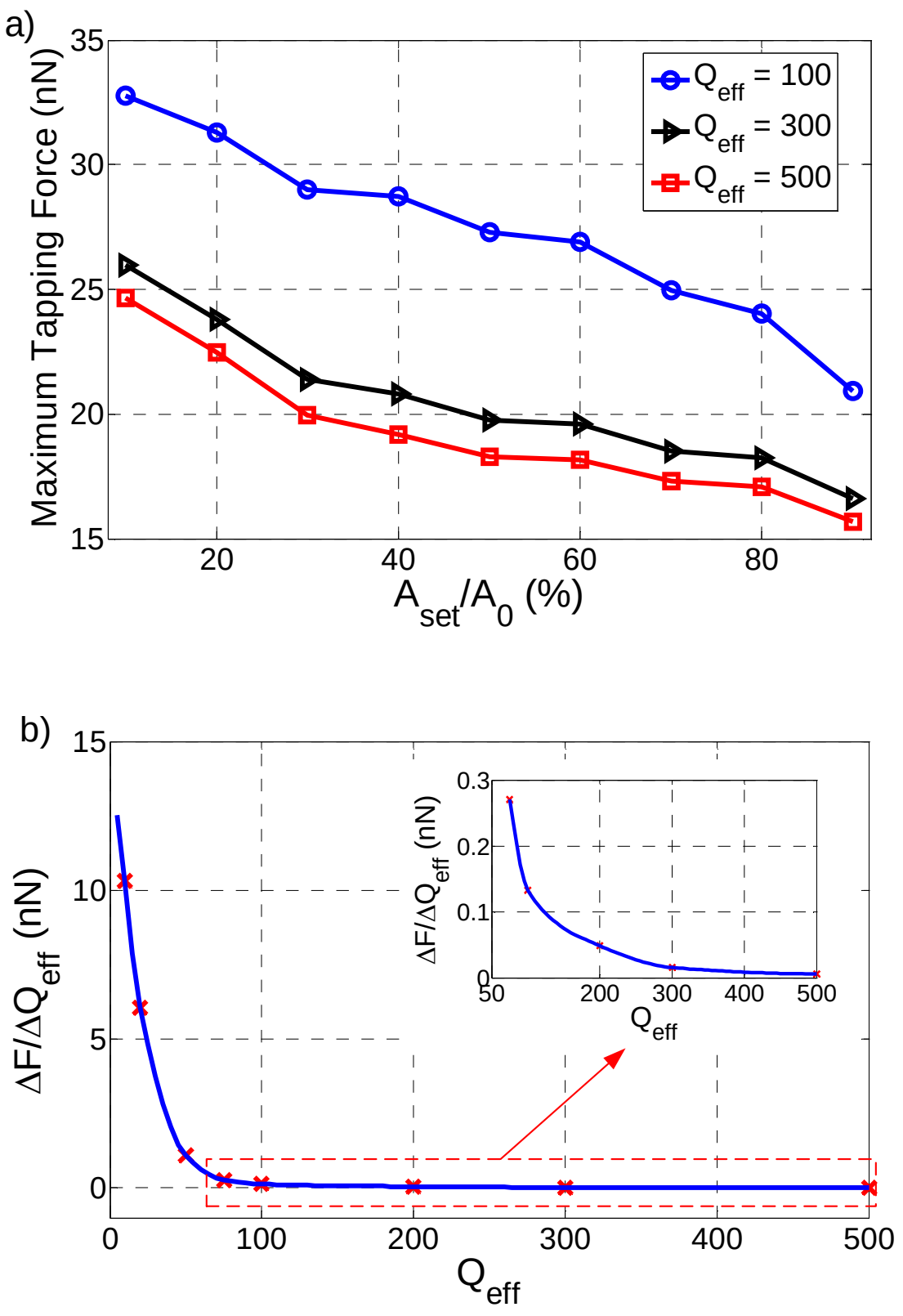

**Figure 5** (a) Maximum tapping forces as a function of effective quality factor for different set amplitudes. (b) The rate of change in tapping forces as a function of effective quality factor for $A_{set}/A_0$ = 70%. The elastic moduli of the sample and the substrate are taken as 179 GPa.

Increasing the Q factor, especially when operating in liquid, has the benefits of decreasing the tapping forces and improving the sensitivity of probe such that soft samples can be traced better. However, if the sample surface is stiffer and the native Q factor of the cantilever probe is already high, which is typically the case when scanning in air, the probe sensitivity has a limited influence, but its slow transient response adversely affect the image quality. The bandwidth of the cantilever that is how fast it responds to the changes in surface profile is inversely proportional to its Q factor. As a result, increasing the Q factor limits the maximum achievable scan speed [11, 12]. In order to show the effect of Q factor of a cantilever on its bandwidth and in return to the maximum achievable scan speed, we present the simulation results of scanning a 100 nm step for three different values of $Q_{eff}$ = 10, 400, and 1000 (figure 6). As the $Q_{eff}$ is increased, the response of the probe to the rapid changes in surface profile becomes slower and it cannot trace the input profile well. As shown in figure 6, it takes longer time for the cantilever to return to the set amplitude when a downward step is encountered (figure 6).

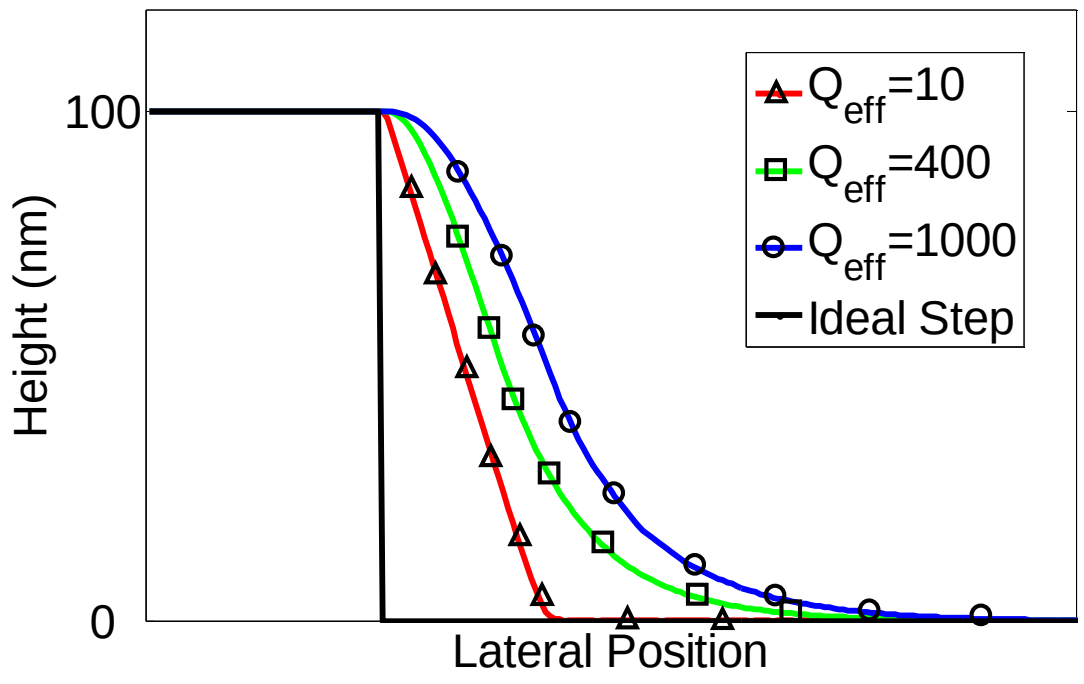

**Figure 6** Scanning a downward step using a cantilever probe having different values of effective quality factor. The elastic moduli of the sample and the substrate are taken as 70 GPa and 179 GPa, respectively.

Therefore, there is a trade-off between the maximum achievable scan speed and the tapping forces applied to the sample when scanning under Q control. Higher Q values promote lower tapping forces while lower ones result in an increased scanning speed. In order to show this trade-off more quantitatively, we performed simulations of scanning in tapping mode AFM for different values of $A_{set}$ and scan speeds. We have defined an error measure to evaluate and compare the scan performances under different scan settings. This measure is based on the positional error between the measured (output) and desired (input) scan profiles. We first define $e_x$ as the absolute value of the positional difference between the measured and actual heights of the sample surface at a lateral position x along a scan line. Then, the total scan error, $e_s$, is calculated by integrating the positional error $e_x$ over the scan-line. The magnitude of the scan error is zero if the resultant scan profile is exactly matches the actual surface profile. In the case of scanning calibration steps with a constant height of $h_s$ (figure 7), the total scan error is calculated by first integrating the positional error $e_x$ over a full step width $w_s$ and then normalizing the sum by the area under the step. This normalization makes the scan error invariant of the scan speed so that it can be used for comparing the results of different scan speeds. Hence, the scan error $e_s$ is calculated as

$$e_s = \frac{\int_0^{w_s} |e_x| dx}{w_s h_s} \tag{7}$$

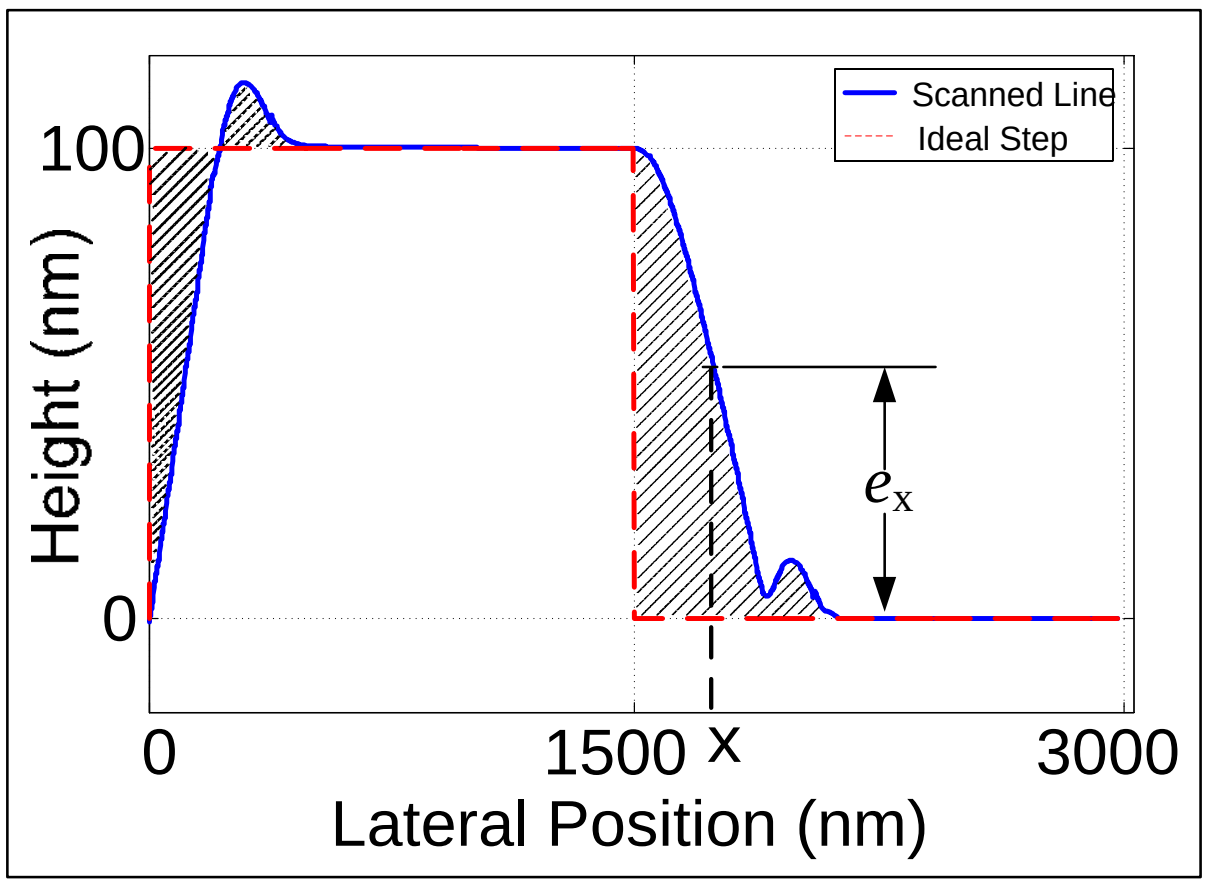

**Figure 7** Scan profile (blue solid line) of an ideal step (red dashed line) with a height of $h_s$ = 100 nm and width of $w_s$ = 3 μm.

In figure 8, we present the iso-error curves obtained by scanning a 100 nm step under Q control for $Q_{eff}$ = 100 and 500. It is observed that the scan error increases as the scan speed and $A_{set}$ increase. Also, increasing the Q factor of the cantilever causes higher scan errors. However, recall that tapping forces are reduced as the Q factor is increased (figure 5a). The influence of $A_{set}$ on this trade-off becomes more prominent when operating at higher scan speeds (iso-error lines become more inclined at higher scan speeds in figure 8).

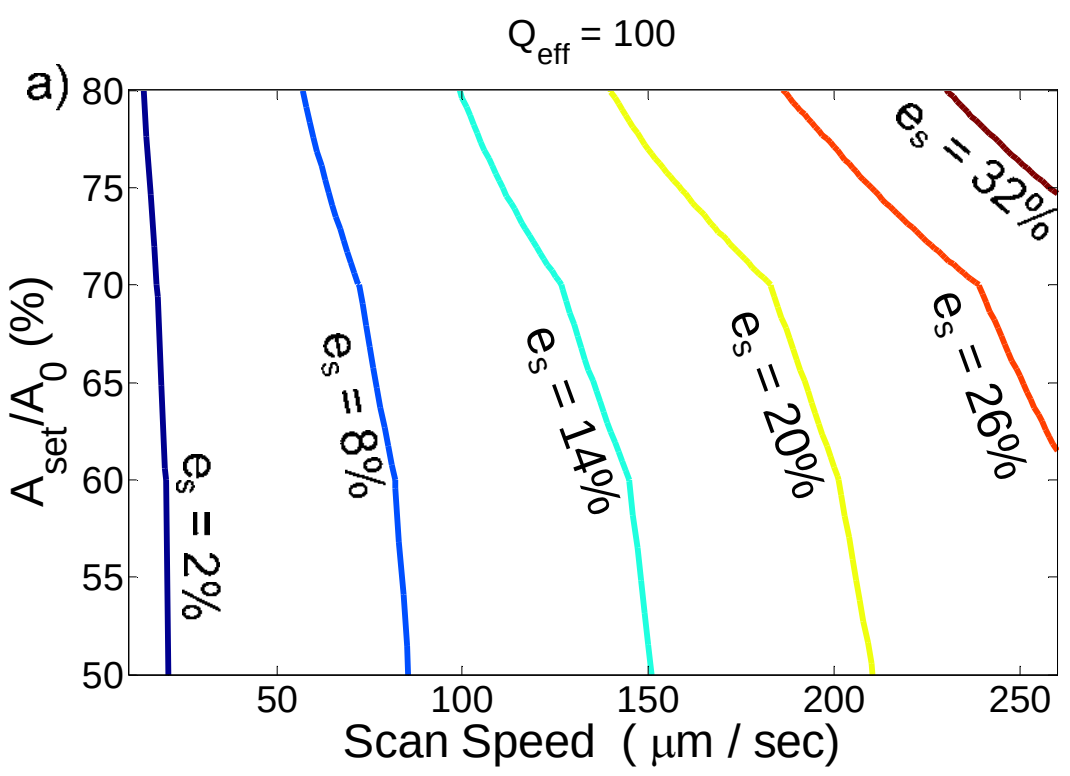

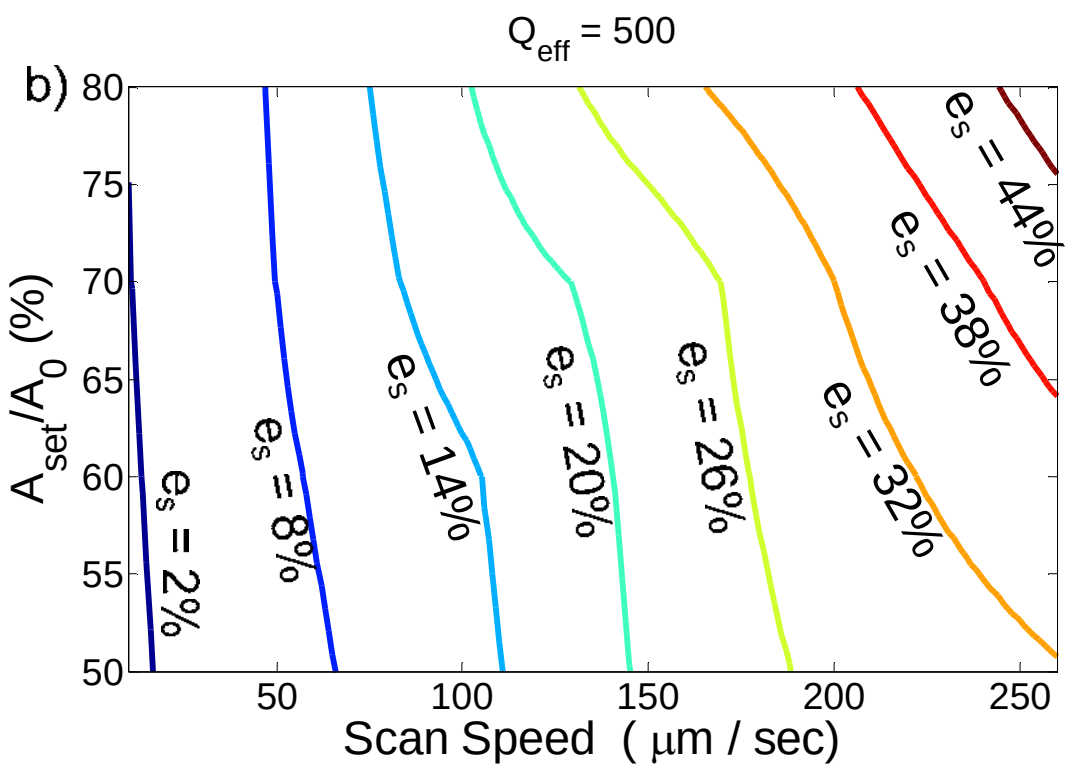

**Figure 8** Iso-error curves for the scan results of a 100 nm step for (a) $Q_{eff} = 100$ and (b) $Q_{eff} = 500$.

In figure 9, we compare the performance of standard Q control and AQC using iso-error lines obtained by numerical simulation of scanning 100 nm steps. For a fixed scan error of $e_s$ = 20% and $A_{set}/A_0$ = 70%, one can easily reach higher scan speeds using AQC for $Q_{eff} = 300$. As shown in the figure, one can also reach the similar scan speeds by setting the Q factor to $Q_{eff} = 5$ in standard Q control, but this causes an increase of approximately nine times in the maximum tapping force applied to the sample when it is compared to case of AQC for $Q_{eff} = 300$.

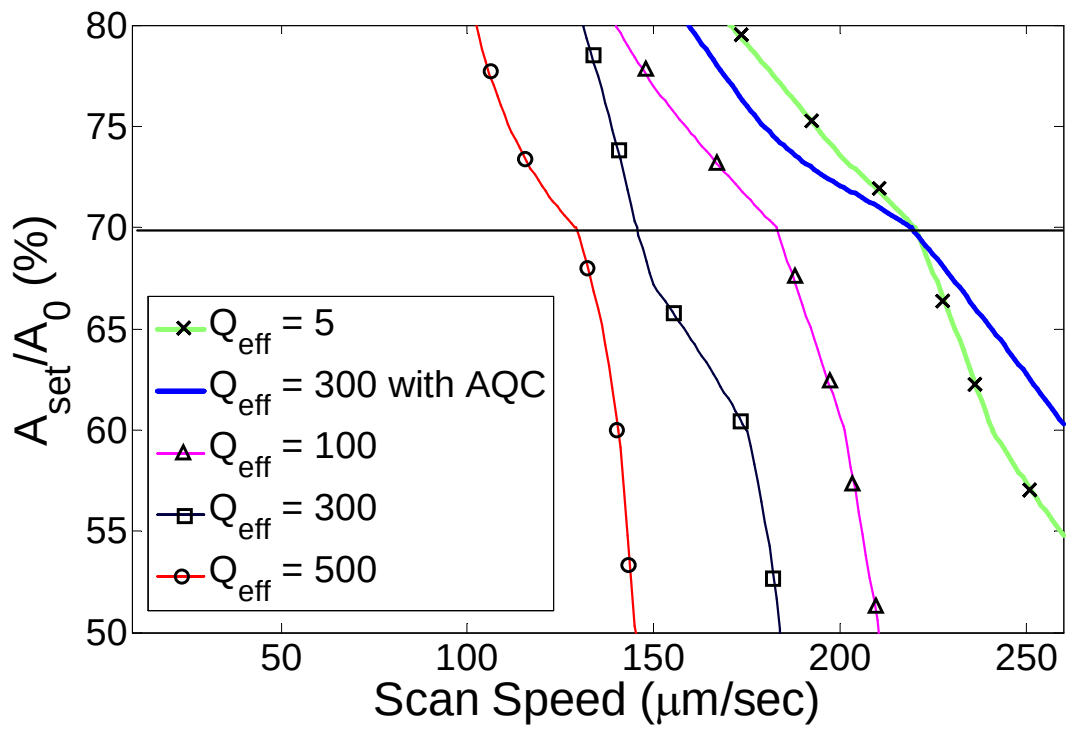

**Figure 9** Comparison of standard Q control and AQC using iso-error curves for the scan error of $e_s$= 20%.

## 6. Experiments

While the tapping forces are reduced as Q factor is increased, the transient response of the probe becomes slower and hence the error signal saturates for a longer period of time, limiting the maximum achievable speed. For example, when scanning steps with a constant height, the error saturation occurs at the beginning and also at the end of the step. When an upward step is encountered, the probe tip suddenly sticks to the surface, the oscillation amplitude reduces to zero and the error signal ($A_{set}$ – A) saturates at $A_{set}$ (see the error signal between the lateral positions p1 and p2 in figure 10). When a sharp downward step is encountered, the oscillation amplitude of the probe reaches to its free air value and the error signal this time saturates at $A_{set} - A_0$ (see the error signal between the lateral positions p3 and p4 in figure 10). Choosing a high value for $A_{set}$ reduces the saturation problem at the beginning of the step (since the magnitude of the error signal is high, the controller responds more rapidly) and also reduces the tapping forces (figure 5a), but amplifies the saturation problem at the end of the step. These saturations result in an inclined profile at the entrance and exit of the step (see scan profiles in figure 10).

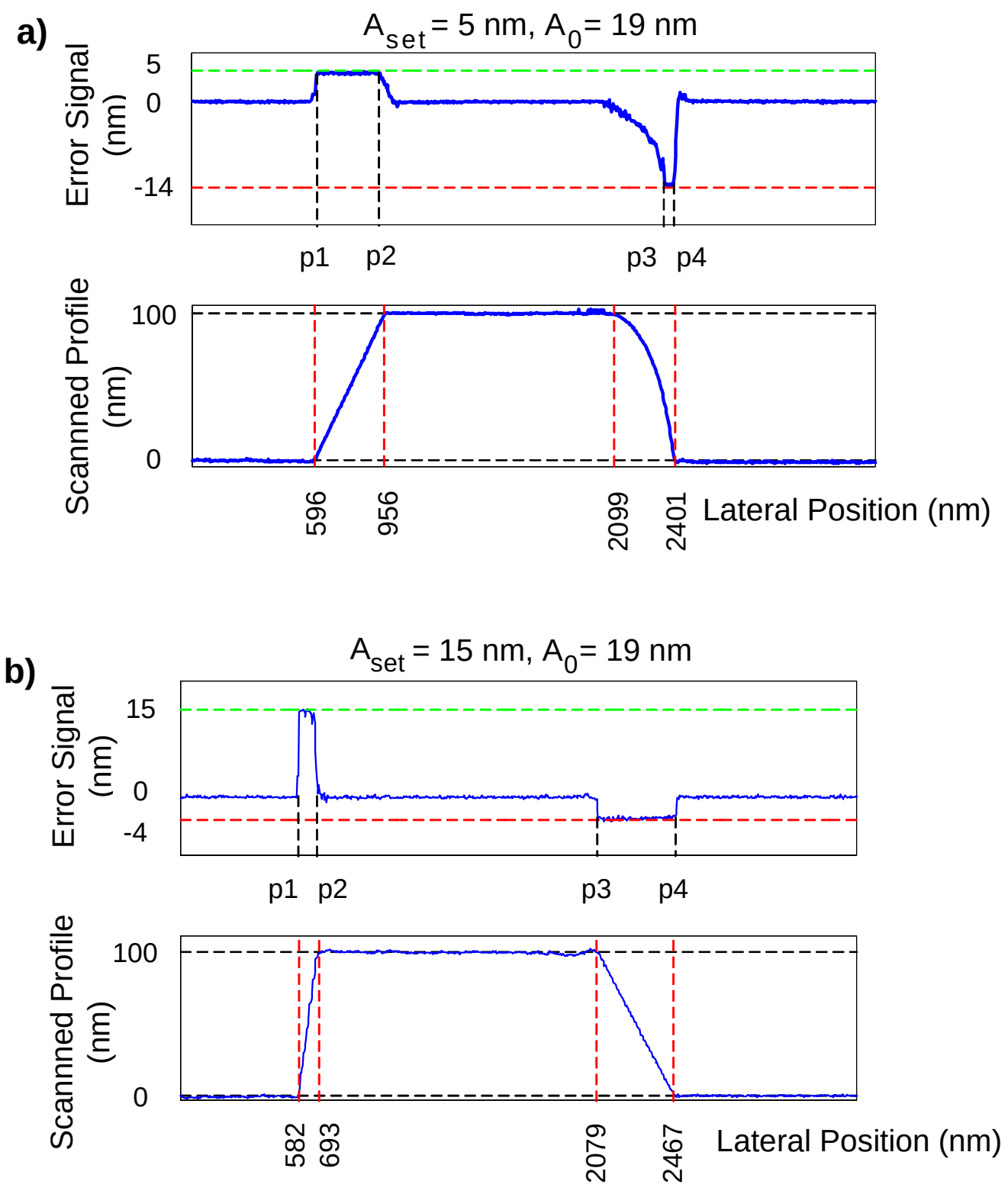

**Figure 10** The experimental results of scanning 100 nm calibration steps at two different settings of $A_{set}$ = 5 nm and 15 nm ($A_0$ = 19 nm). The lateral positions where the error saturation occurs are p1 to p2 and p3 to p4.

The error saturation is also adversely affected by the increase in scan speed. This, in fact, limits the maximum achievable scan speed. In figures 11a and 11c, we present the experimental results of scanning 100 nm calibration steps using conventional PI scan controller for two different scan speeds of 2 µm/sec and 10 µm/sec, respectively. The other scan parameters are kept constant during the experiments and hence the elapsed time during error saturations are the same at both scan speeds. However, the lateral distance traveled during the error saturated period is different, resulting in a less accurate scan profile for the faster scan (figure 11c). Therefore, the error saturation is a major problem limiting the maximum achievable scan speed when scanning surfaces having sharp changes in topography. The same scanning experiments are simulated numerically and the results are presented in figures 11b and 11d. The scan parameters used in the simulations are matched to the ones used in the physical experiments. The saturation problem is also observed in the numerical simulations and the scan profiles obtained through numerical simulations show a good agreement with the experimental ones.

The AQC reduces the error saturation problem leading to an increase in scan speeds with minimal increase in tapping forces (figure 11e). As an alternative, the integral gain constant of the PI controller can be increased to overcome the saturation problem [23]. However, there is an upper limit for its value. A higher controller gain magnifies the noise in the measurement (i.e. cantilever deflection signal) and significantly reduces the quality of the resultant image.

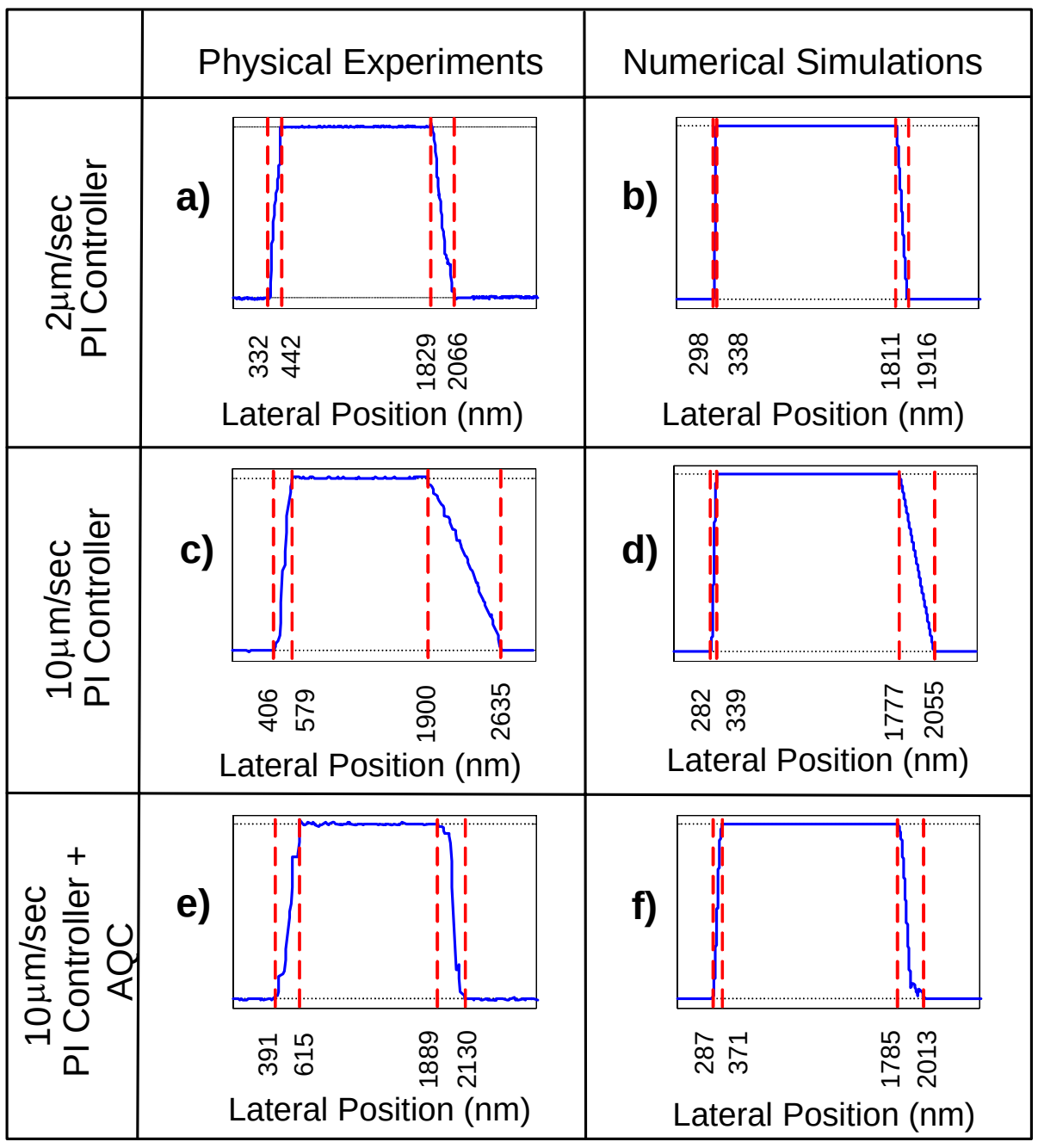

**Figure 11** The results of the experimental and numerical scans without AQC for the scan speeds of 2 µm/sec (a and b) and 10 µm/sec (c and d) and with AQC for the scan speed of 10 µm/sec (e and f). More experimental results comparing standard and adaptive Q control are available in our earlier publication [13].

As can be seen from the figure 11e, the cantilever probe tracks the input profile better in AQC, resulting in a sharper image. In AQC, the Q factor of the probe is set to an initial value as in standard Q control, but then modified on the fly during scanning when necessary. For example, if an error saturation is detected when scanning a downward step, AQC automatically increases the gain G (and hence the Q factor of the probe). This saturation occurs when $A > A_{threshold}$, where $A_{threshold}$ is a threshold value close to $A_0$. An increase in the Q factor causes an increase in the vibration amplitude A as well as the magnitude of the error signal (figure 12), which results in a faster response of the z-actuator moving the sample. As a result, the adverse effects of the error saturation on the output profile are significantly reduced using AQC. Moreover, since the free-air saturation problem is suppressed using AQC, one can use a higher $A_{set}$ to reduce the tapping forces as well as the error saturation at the beginning of the step as discussed earlier.

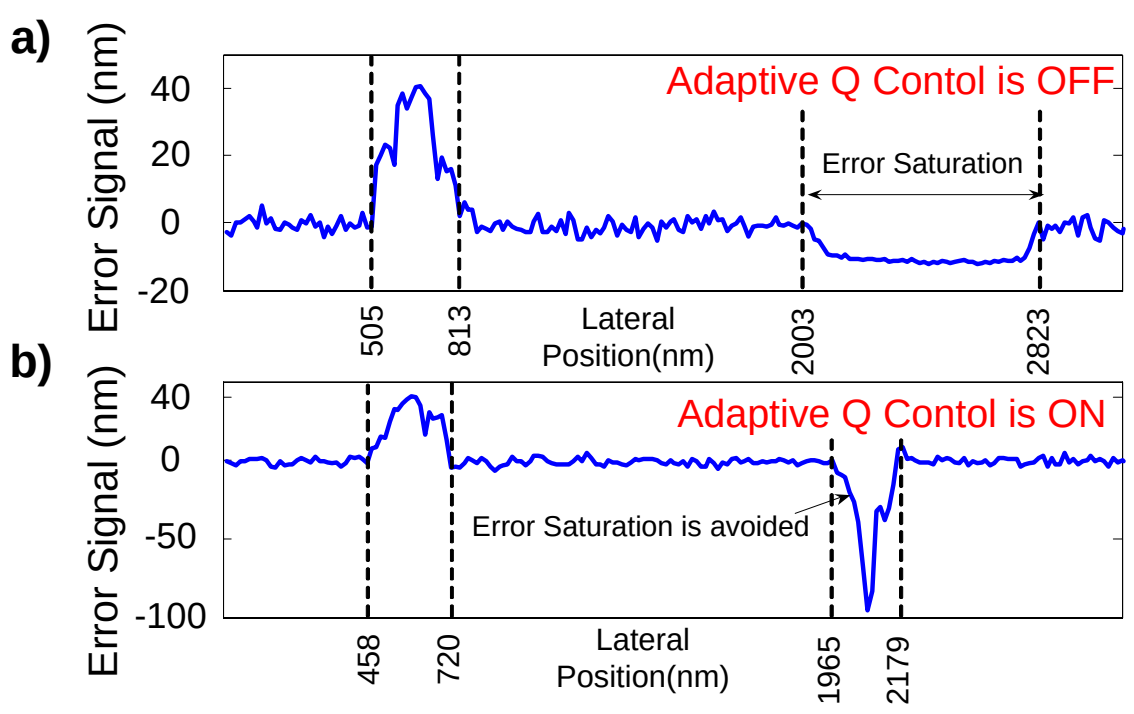

**Figure 12** The error signal recorded during the scanning of 100 nm calibration steps (scan speed = 10 µm/sec) when the adaptive Q controller is turned off and on. The error saturation observed in (a) is avoided using AQC (b).

## 7. Discussion and Conclusions

The trade off in setting the Q factor of a cantilever probe was investigated numerically and the performance of standard Q control was quantitatively compared to that of AQC using a new error measure. This error measure is based on the physical dimensions of the surface being scanned only and for someone interested in the material properties of the sample, it does not help to compare different scans. The results of the numerical simulations were validated through scan experiments performed in a physical set-up. Although numerical simulations have been performed in the past to investigate the cantilever dynamics and tip-sample interactions, only a few recent studies have focused on the simulation of whole scan process [14]. Since it is almost impossible to repeat a scan experiment under the same exact conditions in real world settings due to variations in humidity, electrical noise, ground vibrations, temperature, etc, developing numerical simulations is important. Moreover, even if the experimental conditions are fixed, the probe tip or the sample surface may get damaged in time, affecting the scan results adversely. All these factors make it impossible to quantitatively compare the results of experimental scans under different scan settings. On the other hand, we, for example, easily constructed the performance lines of constant error for different scan settings using numerical simulations to compare the performance of standard Q control with AQC in our study. In addition, we readily accessed all the outputs of the simulations, in which some are not directly measurable in physical experiments (e.g. magnitude of maximum tapping forces shown in figure 5).

Moreover, one can also easily run “what-if” scenarios in a simulation world to investigate the scan performance not only for different settings of scan parameters, but also for different scan environments such as liquid and air. For example, we investigated the effect of stiffness constant and Q factor of a cantilever on the scan profile when scanning a soft sample lying on a stiffer substrate (figure 4). We showed that scanning in attractive regime using soft cantilevers having high $Q_{eff}$ results in a better image quality. If a soft sample (relative to the substrate surface) is scanned in liquid where the Q factor drops significantly, the probe tip cannot accurately trace the surface profile and the height of the sample surface is measured less than its actual value. Moreover, there is a risk of damaging the sample due to high tapping forces at low values of Q (figure 5a). To overcome these problems, the Q factor of the cantilever is increased when scanning in liquid. On the other hand, when scanning in air, the native Q factor of the cantilever is typically high and increasing it further does not significantly reduce the tapping forces. Our simulation results show that the magnitude of tapping forces is high at low values of $Q_{eff}$ and decreases as the $Q_{eff}$ is increased while the rate of drop is not significant at higher values (figure 5b). Moreover, the slow transient response of the cantilever at high values of $Q_{eff}$ limits the maximum achievable scan speed. Hence, reducing the Q factor increases the scan speed, but it also increases the magnitude of tapping forces, which may cause damage to the sample and/or the probe tip. However, it is possible to achieve higher scan speeds using AQC without causing an increase in the tapping forces (figure 9). We showed that error saturation is the major problem limiting the scan speed in tapping mode AFM under standard Q control and AQC solves this problem (figure 12). In AQC, if there is a tendency towards saturation in the error signal due to the rapid variations in surface topography, the controller changes the Q factor of the probe instantaneously to avoid the saturation problem. Moreover, since the error saturation problem is suppressed, one can set higher values of $A_{set}$ to reduce the tapping forces as well (see the relation in figure 5a).